\documentstyle[epsf,prd,aps,floats]{revtex}  
   
\newcommand{\be}{\begin{equation}}    
\newcommand{\ee}{\end{equation}}    
\newcommand{\bea}{\begin{eqnarray}}    
\newcommand{\eea}{\end{eqnarray}}

\begin{document}     
\twocolumn[\hsize\textwidth\columnwidth\hsize\csname@twocolumnfalse\endcsname     
\author{Jo\~ao Magueijo}    
\date{\today}  
\address{Theoretical Physics, The Blackett Laboratory,  
Imperial College, Prince Consort Road, London, SW7 2BZ, U.K.}  
\title{Stars and black holes in varying speed of light theories}   
  
\maketitle     
\begin{abstract}  
We investigate spherically symmetric solutions to a recently proposed 
covariant and locally Lorentz-invariant varying speed of light theory.  
We find the metrics and variations in $c$ associated with the  
counterpart of black holes, the outside of a star, and  stellar collapse. 
The remarkable novelty is that $c$ goes to zero or infinity (depending  
on parameter signs) at the horizon. We show how this implies that,   
with appropriate parameters, observers are prevented from entering the 
horizon. Concomitantly stellar collapse must end in a ``Schwarzchild  
radius'' remnant.  We then find formulae for gravitational light deflection,  
gravitational redshift, radar echo delay, and the precession of the  
perihelion of Mercury, highlighting how these may differ distinctly from 
their Einstein counterparts but still evade experimental constraints. 
The main tell-tale signature of this theory is the prediction of the 
observation of a different value for the fine structure constant, 
$\alpha$, in spectral lines formed in the surface of stars. We close
by mentioning a variety of new classical and quantum effects near 
stars, such as aging gradients and  particle production.
\end{abstract}    
\pacs{PACS Numbers: *** }  
]

\renewcommand{\thefootnote}{\arabic{footnote}}  
\setcounter{footnote}{0}

\section{Introduction}     
The possibility that the speed of light $c$ might vary has  
recently attracted considerable attention \cite{mof1,am,b,bm1,bm2,bm3,bm4,mof2,cly1,cly2,cly3,drum,av1,av2,harko,kir,alex,bass,li,elec,jac}. 
Most notably, in a cosmological setting, temporal variations 
in $c$ have been shown to solve the so-called cosmological puzzles -  
the horizon, flatness, and Lambda problems of Big-Bang cosmology. 
At a more conceptual level it is clear that varying speed of light  
(VSL) theories require extreme departures from the standard framework 
of physics, since they contradict the leading postulate behind relativity 
and Lorentz-invariance.  A number of alternative implementations 
for VSL have been discussed, involving either hard \cite{am}  
or soft \cite{mof1} breaking of Lorentz invariance.  
 
In a recent paper \cite{li} 
it was shown that contrary to popular belief it is  
possible to set up covariant and  
locally Lorentz invariant VSL theories, as long as these concepts 
are subject to very minimal generalizations. As a matter of fact 
the necessary generalizations glean from the usual definitions  
all that is operationally meaningful, in the sense that   
the aspects they  preserve are exactly those which  
can be the outcome of experiment. Such a formulation 
arguably provides the most conservative VSL theory one may set up. 
It is found that in such theories the local value of $c$ is determined via 
a differential equation, containing as source terms the cosmological 
constant and the matter Lagrangian. 
 
Naturally in such theories $c$ varies not only in time (over cosmological  
time scales) but also in space, once the inhomogeneity of the Universe 
is taken into account \cite{otoole}. In the simplest case one 
should investigate such a phenomenon by seeking static and 
spherically symmetric solutions. Such is the purpose of this paper. 
We investigate VSL solutions representing the counterpart of black holes,  
the exterior of a star, and stellar collapse. It should be stressed 
that it is such solutions, not the cosmological ones, that 
bear relevance to many experimental tests (a point  
entirely missed by \cite{elec}). 
 
In Section~\ref{summary} we start by reviewing the key aspects 
of the theory proposed in \cite{li}. Then in Section~\ref{bh} 
we consider static spherically symmetric solutions, both in 
isotropic and radial coordinates. We find  the limit under which 
the Schwarzchild solution is still a solution of our theory, and note 
that $c$ goes to zero or infinity at the horizon. 
We also find the most general solution, which is similar 
to the solution found in \cite{dick2}; however the relationship 
between the various parameters in \cite{dick2} 
is new. In all of these solutions we 
find that $c$ must go to zero or infinity at the horizon. This is 
not accidental, and  
in Section~\ref{proof} we sketch a proof showing why this is 
generally the case.  
 
The last result has two very significant implications. The first  
is discussed in Section~\ref{inaccess}, and corresponds to the 
naive expectation that if $c$ goes to zero fast enough at the 
horizon then no observer can actually reach it.   Indeed 
$c$ still acts as a local speed limit. This insight proves 
to be true, even when a number of complications are taken into 
account. Firstly the field $c$ may also act as a gravitational 
field, pushing free-falling particles off geodesics, accelerating 
or braking them. Secondly, as $c$ changes so do all fine structure 
constants, and also the time rates of the interactions they promote. 
One should attach the definition of time to these rates, and examine 
the problem of an observer falling into a black hole from the point of 
view of the number of ticks of such ``interaction time''. We find that 
when all this is taken into account, there is 
still a large region of parameter space for which reaching the  horizon  
requires infinite free-falling time. VSL black holes are therefore 
not covered by an ``horizon'' but instead, the horizon represents 
an edge of space-time to be put on the same level as the asymptotic 
spatial infinity. This has the implication that the singularity 
may be excised from the manifold - we conjecture that perhaps one 
can get rid of all singularities in a similar way in VSL theories. 
 
Another interesting result is described in Section~\ref{col}: stellar 
collapse may take infinite interaction time when viewed by an observer 
on the surface of the collapsing star. This follows directly from the 
above considerations, and implies that the end point of stellar 
collapse must be a Schwarzchild remnant.  As this is formed the speed 
of light goes to zero for all points inside the star, thereby 
freezing all processes and preventing the formation of a singularity. 
 
The final part of this paper, contained in Section~\ref{phys}, 
is devoted to possible experimental tests for this theory based  
upon solar system gravitational physics. We concentrate on the classical 
tests of GR (General Relativity),  
leaving to a future publication the analysis of more recent 
(but more complex) experiments, such as the binary pulsar  PSR
1913 + 16
\cite{will}. 
We examine the effects 
of VSL upon the orbits of planets,  gravitational 
light deflection, and the radar echo time-delay. The real novelty 
is, however, the effects upon the spectral lines  formed 
at the surface of stars, 
for which our theory predicts a fine structure different from  
laboratory measurements. We find that it is possible to reproduce 
all the standard GR tests, and still have 
a non-negligible spectral effect. The application of techniques 
similar to the ones developed by Webb et al \cite{webb} 
should put this theory to the test.  
 
We conclude with a brief qualitative discussion of an assortment of  
exotic new phenomena expected in the vicinity of very massive 
stars in  VSL theories.  
  
\section{Summary of the theory}\label{summary} 
We first summarise the covariant and locally Lorentz invariant 
VSL theory proposed in \cite{li}. In this theory  
the speed of light plays 3 distinct roles 
(corresponding to independent aspects of the theory) 
parameterized by numbers $q,\kappa,a,b$, and $\beta$. 
 
At its most innocuous, VSL is nothing but a theory predicting changing  
fine structure constants $\alpha_i=g_i^2/(\hbar c)$ (in which 
$i$ label the various interactions, and $g_i$ are charges), 
with fixed ratios $\alpha_i/\alpha_j$. 
Choosing units such that changes are attributed  
primarily to $c$ is useful simply because they lead to a simpler picture. 
A fixed-$c$ dual theory may be obtained by a change of units, 
but the ensuing local dynamics is then rather contrived.  
Also, important global features 
may be missed in fixed-$c$ units (e.g. the trans-eternal  
regions, or the black hole edges discussed in \cite{li}). In \cite{li} we then  
required that the matter Lagrangian should  
not depend on $c$; this fact alone fixes the scaling with $c$ of all  
Lagrangian parameters up to the $\hbar(c)$ dependence. In particular 
particle rest energies scale like $E_0\propto \hbar c$, 
and all gauge charges like $g_i\propto \hbar c$. Taking $\hbar c \propto c^q$ 
we then have $\alpha_i\propto g_i\propto \hbar c \propto c^q$. 
In summary,  $c$'s first role is to parameterize changes in all ``constants''  
in minimal changing $\alpha$ theories for which the Lagrangian itself  
is required to remain invariant. 
 
One must then endow $c$ with its own action, and note that $c$ 
appears in the gravitational Lagrangian as part of a conversion 
factor between curvature and energy density. As pointed out in 
\cite{li} the definition of $c$  
in terms of a field, its dynamics, and its coupling to gravity 
and matter may be defined in many different ways. In the 
simplest $\psi=\log(c/c_0)$, and 
\be\label{S21}     
S= \int d^4x \sqrt{-g}{\left( e^{a\psi}(  
R +{\cal L}_{\psi})  
+{ 16\pi G\over c_0^4}e^{b\psi}{\cal L}_m \right)}  
\ee  
where we shall not impose $a-b=4$ (and we have set  
$\Lambda=0$). 
The simplest dynamics for $\psi$ derives from:  
\be  
{\cal L}_{\psi}=-\kappa(\psi) \nabla_\mu\psi\nabla^\mu\psi   
\ee  
Hence the second aspect of $c$, as a dynamical field and coupling 
constant, is parameterized by $a,b$ and $\kappa$.  
In \cite{li} it was shown how this aspect of the theory allows for analogies 
with some string theories to be made \cite{damour}. 
 
Thirdly,  
the theory proposed in \cite{li} is covariant and locally Lorentz invariant, 
in a generalized sense which accommodates a varying $c$. The 
generalization is trivial; and essentially amounts to the 
use of an $x^0$ coordinate in all differential geometry formulae. 
The only significant difference 
is that if $c$ varies, local measurements of space $dx$ and time  
$dt$ do not generally lead to closed forms (ie. $d^2 t\neq 0$ 
or $d^2 x\neq0$), leading to a fibre bundle structure where 
usually one finds a tangent bundle. 
However they admit integrating factors, so that  
$dt\psi^\beta$ and $d x\psi^{\beta -1}$ are closed forms. 
Hence $c$ appears in a third role, as a conversion factor between 
space and time, and as an integrating factor defining the change  
of units which would convert the theory into a fixed $c$ standard 
covariant and locally Lorentz invariant theory. The parameter 
$\beta$ needs not be related to any other parameters, but we considered 
the cases $\beta=3-q/2$ and $\beta=1-q/2$. 
 
The equations for such a theory are: 
\bea \label{einstein21}     
G_{\mu\nu}&=&{8\pi G\over c_0^4 e^{(a-b)\psi}}     
T_{\mu\nu}+ \kappa{\left(\nabla_\mu\psi \nabla_\nu\psi     
-{1\over 2}g_{\mu\nu}\nabla_\delta\psi \nabla^\delta\psi   
\right)}   \nonumber \\  
&&+e^{-a\psi}(\nabla_\mu\nabla_\nu e^{a\psi}  
-g_{\mu\nu}\Box e^{a \psi})    
\eea  
and  
\bea\label{psi21}   
\Box \psi&&  +a\nabla_\mu\psi \nabla^\mu\psi\nonumber\\  
&&={8\pi G\over c_0^4 e^{(a-b)\psi}(2\kappa + 3a^2)}  
(a T -2b  
{\cal L}_m )  
\eea  
A change of units rephrases these theories as 
Brans-Dicke theories \cite{dick2} 
only when $b+q=0$ and $\beta=1-q/2$. However there is a formal 
analogy between action (\ref{S21})  and  
Brans-Dicke theory in the Jordan frame, established with 
the following identifications: 
\bea 
\phi_{bd}&=&e^{a\psi}\label{bdanal1}\\ 
\omega_{bd}&=&{\kappa\over a^2}\label{bdanal2}\\ 
T^{bd}_{\mu\nu}&=&e^{b\psi}T_{\mu\nu}\label{bdanal3} 
\eea 
The analogy is always valid in vacuum, but breaks down 
when $b\neq 0$ inside matter distributions. Indeed $T^{bd}_{\mu\nu}$ 
then depends on $\phi_{bd}$ {\it in the Jordan frame}. Bearing this 
in mind, we shall make use of this analogy for reading off solutions  
from \cite{dick2}. However careful rederivation will be required to account 
for  novelties induced by  $b\neq 0$.  
 
A further analogy with scalar-tensor theories arises from the  
conformal equivalence of various $\{a,b,\kappa\}$ theories. 
Conformal transformations do not change the speed of light, 
mapping VSL theories into VSL theories; but the gravitational action is 
modified leading to different values for $a$, $b$, and $\kappa$.  
This may be used to 
simplify the dynamics, in particular reducing it to Brans-Dicke 
dynamics. The relevant transformations are 
spelled out in Appendix II, where a set of results 
is derived which may then be used to provide alternative derivations 
for many results in the main body of this paper.  
 
However, as stressed 
in Appendix II, the frame realizing Brans-Dicke dynamics can only be 
achieved with very restricted forms of matter. In particular,  
one must require that ${\cal L}_m$ be homogeneous in the metric;  
clearly far from 
true in general. In the particular case in which we only consider 
classical point particles the Lagrangian takes the form: 
\be\label{partS} 
S=-{E_0\over 2\alpha}\int d\lambda [-g_{\mu\nu}\dot x^\mu\dot x^\nu]^\alpha 
\ee 
in which $\alpha$ can a priori be any number. In metric theories 
of gravity the value of $\alpha$ is irrelevant, because $u^2$ 
(with $u=\dot x$) is a constant. One usually takes $\alpha=1/2$, 
so that the action becomes the length of the world-line. 
The value of $\alpha$ is however physically  
relevant if $b\neq 0$ \cite{li}, and the results in this paper 
do depend on $\alpha$. Arguments for $\alpha=1$ were put forward 
in \cite{li}, and we shall adopt this assumption in the main body of 
this paper. This implies that for classical point particles  
${\cal L}_m=-\rho/2$, with $\rho$ the energy density.  
Hence ${\cal L}_m$ is homogeneous degree 1 in the metric.

For general forms of matter, minimal coupling, that is, the requirement that  
${\cal L}_m$ does not depend on $c$, is not conformally invariant. 
Therefore a conformal frame (and so a set of $a$ and $b$) is picked for 
its simplicity in describing non-gravitational physics (a point 
clearly made in \cite{qui}). This renders the construction described 
in Appendix II a useful mathematical tool, but with limited physical 
meaning, except when the generality of ${\cal L}_m$ can be swept under 
the carpet.

\section{Vacuum spherically symmetric solutions} \label{bh} 
Let us consider static spherically symmetric (SSS) solutions.  
We shall work with both radial coordinates: 
\be 
ds^2=-B d\xi^2 + A dr^2 + r^2 d\Omega^2 
\ee 
(where $d\Omega^2=d\theta^2+\sin^2\theta d\phi^2$) and with isotropic 
coordinates: 
\be 
ds^2=- F d\xi^2 + G (d\rho^2 +\rho^2 d\Omega^2) 
\ee 
Recall that as in \cite{li} the usual tools of differential 
geometry are unaffected under the condition that an  $x^0$-type 
of coordinate is used, here denoted by $d\xi=cdt$.

\subsection{The VSL Schwarzchild solution}  
The simplest SSS solution to VSL is the Schwarzchild solution:  
\be\label{schw} 
ds^2=-{\left(1-{2Gm\over c_\infty^2 r}\right)}d\xi^2   
+{dr^2\over 1-{2Gm\over c_\infty^2 r}}+r^2d\Omega  
\ee  
This is valid whenever the field $\psi$ does not gravitate,  
eg. in the bimetric theory discussed in the Appendix of \cite{li}  
(a case developed further in Appendix I).  
This is also true in the theory  described above   
in the limit $\kappa,a\rightarrow 0$. We may then have
$\kappa/a$ and $\kappa/b$ finite, or  $\kappa/a$ and $b$ 
finite if $\kappa/a\gg b$. As we will see the latter case distinguishes
itself by predicting non-geodesic motion (but no corrections
to the metric) while the former predicts geodesic motion.

The horizon is at $r_h=2Gm/ (c_\infty^2r)$, and the mass  
$m$ is identified by comparing $g_{00}$ and the weak field 
solution to this theory. Later we shall see that $m$ need 
not be the Keplerian mass, if $b\neq 0$ (the case in which
planets do not follow geodesics).
 
Integrating (\ref{psi21}) with  metric (\ref{schw})  
leads to the exact solution:  
\be 
\psi={b-a\over 2\kappa}\log{\left(1-{2Gm\over c_\infty^2 r}  
\right)}  
\ee 
in which the factor $(b-a)/(2\kappa)$ can be found using the
weak field limit. Hence:
\be\label{cschw} 
c=c_\infty{\left(1-{2Gm\over c_\infty^2 r}\right)}^{b-a\over 2\kappa}  
\ee  
We see that the speed of light goes to either zero or infinity at the  
horizon depending on the couplings,  a property we shall 
prove in general in Section~\ref{proof}. 

Physically the effect of the coupling parameters' signs and
relative magnitudes is as follows.
Let $\kappa>0$ so that the energy in the VSL field $\psi$ is
positive (but negligible, since $\kappa\rightarrow 0$). The field
$\psi$ is then driven by direct couplings to matter and to gravity, 
with strengths proportional
to the couplings $b$ and $a$ respectively (cf. Eqn (54) of \cite{li}).
If both $b$ and $a$ are positive 
the first coupling drives $c$ to decrease close to matter concentrations,
the second to increase. If $b=a$ (such as in the case
of the dilaton coupling at tree level, as discussed in \cite{li})
the speed of light does not change
near matter concentrations. If $b>a$ light slows down
close to massive bodies; if  $b<a$ it speeds up. In either case,
we found that near a black hole's horizon something extreme 
must happen: $c$ must go to either zero or infinity.  
The fact that something extreme must happen is due to the structure 
of space-time, and can be linked to the usual proofs of the no-hair
theorem as we shall see. The choice between the two options
is made by the relative strengths of the $a$ and $b$ couplings, 
and follows whatever trend in $c$ is already present in the weak field
region.

The solution we have just found will be extremely useful in clarifying 
the meaning of more complicated solutions. It preserves the simplicity 
of the Schwarzchild solution while allowing for a variety of non-gravitational 
VSL effects to be present.  
  
\subsection{Brans-Dicke type of solutions} \label{bhbd}
Given the formal analogy in vacuum between VSL theories and Brans-Dicke  
theories, we may  use \cite{dick2} to write the following exact  
solution: 
\bea 
ds^2&=&-F^{2\over \lambda}d\xi^2 + \nonumber \\ 
&& {\left(1+{\rho_0\over \rho}\right)}^4  
F^{2(\lambda-C-1)\over 
\lambda} 
(d\rho^2 +\rho^2 d\Omega^2)\label{bd1}\\ 
c&=&c_0F^{C\over a \lambda}\label{bd2} 
\eea 
with 
\bea 
F&=&{1-\rho_0/\rho\over 1+\rho_0/\rho}\\ 
\lambda^2&=&(C+1)^2-C(1-\kappa C/(2a^2))\label{lamb} 
\eea 
However, the weak field limit imposes a relation between 
$\{C,\lambda,\rho_0\}$ and $\{a,b,\kappa,m\}$ which 
goes beyond the identifications (\ref{bdanal1})-(\ref{bdanal3}).  
This is due to the fact 
that when ``Brans-Dicke'' language is adopted for VSL theories 
the matter Lagrangian now depends on $\phi_{bd}$, when $b\neq 0$ 
(cf. Eqn.~\ref{bdanal3}). 
 
Mimicking the weak field calculation presented in \cite{dick2}, 
we find that (\ref{einstein21}) and (\ref{psi21}) lead to  
\bea 
\psi&=&{a-b\over 3 a^2 + 2\kappa}{2m\over r}\\ 
-g_{00}&=&1-{4m\over r}{2a^2+\kappa -{ba\over 2}\over 3a^2 +2\kappa} 
\eea 
in which recall we have assumed ${\cal L}_m=-\rho/2$  
(cf. Eqn~\ref{partS} and its following discussion). Defining a Poisson mass  
\be\label{bdmass} 
M=2m{2a^2+\kappa -{ba\over 2}\over 3a^2 +2\kappa} 
\ee 
we then have 
\bea 
\psi&=&{a-b\over 2 a^2 + \kappa -ba/2}{M\over r}\label{psi1}\\ 
-g_{00}&=&1-{2M\over r}\label{g001} 
\eea  
We note once more that $M$ need not be the Keplerian mass. 
 
If we now expand (\ref{bd1}) and (\ref{bd2}) we obtain: 
\bea 
\psi&=&-{CM\over ar}\\ 
-g_{00}&=&1-{2M\over r} 
\eea 
with $M=2\rho_0/\lambda$. Comparing with Eqns.~(\ref{psi1}) and 
(\ref{g001}) we gather: 
\be 
C=-{a^2-ba\over 2 a^2 +\kappa - ab/2} 
\ee 
with $\lambda$ to be obtained from (\ref{lamb}). We stress that  
a direct substitution of  Eqn~(\ref{bdanal2})  in the  
Brans-Dicke result \cite{dick2} misses the terms in $b$. 
 
The metric (\ref{bd1}) may be cast into an Eddington-Robertson expansion 
\cite{weinb}: 
\bea 
ds^2&=&-{\left(1-2{M\over \rho} +2\beta {M\over \rho}^2\right)} d\xi^2 
\nonumber\\ 
&&+ {\left(1+2\gamma {M\over \rho}\right)} 
(d\rho^2 +\rho^2 d\Omega^2) 
\eea 
with the PPN parameters $\beta =1$ and  
\be\label{gamma} 
\gamma=C+1={a^2+\kappa+ab/2\over 2 a^2 +\kappa - ab/2} 
\ee 
 
The Schwarzchild limit may be obtained by letting $a,\kappa\rightarrow 
0$, keeping $\kappa/a$ and $b$ finite but with $\kappa/a\gg b$. Then 
$\gamma\approx 1$, $M\approx m$, and the metric  
reduces to Schwarzchild. However the variation in $c$ is 
non-negligible even in this regime: 
\be 
c=c_{\infty}{\left(1-{b-a\over \kappa}{ G m\over c_\infty^2 r}\right)} 
\ee 
Also deviations from geodesic motion, due to $b\neq 0$, may be  
non-negligible. Hence it is possible to introduce two types of new 
VSL effects without modifying the metric, a feature which we shall 
use to solve a variety of problems.


\section{The speed of light must go to zero or infinity at the horizon}  
\label{proof} 
The fact that in the examples above $c$ goes to either zero or infinity  
at the black hole's horizon is far from accidental. It may be 
generally proved by adapting techniques used in proving the 
no-hair theorem \cite{bek1}. 
Here we sketch how such a general proof might proceed, 
taking the particular case of a scalar $c$ (as opposed to a complex 
$c$ undergoing spontaneous symmetry breaking as discussed in \cite{li}, 
or a $c$ derived from a spinorial field).

Let us consider a static, vacuum, not necessarily spherically symmetric 
solution which is asymptotically flat and   
contains an horizon. Let the metric take the form: 
\be 
ds^2=-L d\xi^2+h_{ij}dx^i dx^j 
\ee 
with $L$ and $h_{ij}$ time-independent. 
Let us discuss the problem in terms of $\phi_{bd}=e^{a\psi}\ge 0$, 
which must satisfy 
\be 
{1\over {\sqrt{Lh}}}({\sqrt{Lh}} h^{ij}\phi^{bd}_{,i})_{,j}=0 
\ee 
We first multiply this expression by ${\sqrt{Lh}}$ and integrate over the  
region $\Omega$ bounded by the horizon (where $L$ must go to zero) and  
infinity. Integrating by parts reveals: 
\be\label{integ} 
\int_{\Omega} dx^3 {\sqrt{Lh}} h^{ij}\phi^{bd}_{,i}\phi^{bd}_{,j} 
-\int_{\partial \Omega}{\sqrt{Lh}} \phi^{bd}h^{ij}\phi^{bd}_{,i} 
dS_j=0 
\ee 
The piece of the surface integral corresponding to infinity 
is zero, by virtue of asymptotic flatness.  
 
At this point VSL differs from relativity. In the usual  
GR  
proof one then shows that the integral over the horizon 
must also be zero since $L\rightarrow 0$ there. The only 
escape route is if $\phi_{bd}$ or its gradient blow up at the 
horizon. This is precluded by the requirement that  
the scalar field energy density be finite. Hence the surface 
integral is zero, and  
since the volume integral is semi-positive definite it must be 
zero, so that the identity is satisfied. This implies that  
$\phi_{bd}=0$ everywhere outside the horizon.  
 
Clearly the last part of the argument may break down in VSL,  
because the $\psi$ gravitation may be negligible. Hence its 
divergence at the horizon need not produce  a singularity. 
This is the case in the parameter region which produces a Schwarzchild  
solution. More generally we may define a region in the space 
$\{a,b,\kappa\}$ for which this type of behaviour occurs. 
 
It may also happen that the $\psi$ divergence at the 
``horizon''causes a singularity. For instance \cite{lousto,agn} 
have shown that this happens for $(1+C)/\lambda <2$.  
However such a singular horizon is not a problematic ``naked  
singularity'' in VSL theories because, as we shall see in the next Section,  
for some regions of the theory's couplings information cannot 
flow out of (or into) the singular surface. Hence a singular horizon 
need not have the pathological connotations it has in general 
relativity.  
 
Whatever happens to the metric at the ``horizon'',  
a non-trivial solution for $\phi_{bd}$ always requires 
that the surface integral in (\ref{integ}) diverges at the horizon.  
This implies  
that $c$ must go either to zero or infinity at the horizon. 
 
The generalization of this argument to stationary solutions, to more general 
fields (ie when $c$ is derived from a bosonic invariant  
associated with a fermionic field $\psi$), or in the presence 
of an electromagnetic field, leads to the same 
conclusion.

A word on terminology is in order. We are loosely using the word 
horizon to describe what can in fact be a naked singularity. However, 
in either  case VSL theories predict that such a surface cannot 
be reached, as we shall show in the next Section. Perhaps the 
wording ``black hole edge'' would be more appropriate, since 
such a surface becomes part of the spatial infinity of the  
space-time. However we shall use the expression horizon in what 
follows for simplicity.

\section{The inaccessibility of singularities}\label{inaccess}  
This theorem has the interesting implication that, at least for  
suitable couplings, the horizon, as well as 
the region inside it, are not physically accessible. Naively one  
might expect this to happen if  
$c$ goes to zero sufficiently fast at the horizon. Indeed 
$c$ still acts as a local speed limit, and so $c\rightarrow 0$  
seems to imply that nothing can enter the horizon.  
However two extra complications come into  
the problem: free-falling particles do not generally follow 
geodesics, and interaction rates change (due to changing $\alpha_i$) 
near the black hole. We shall use Schwarzchild VSL 
black holes as an illustration.  
 
\subsection{Free fall into VSL black holes} 
As  pointed out in \cite{li}, $b\neq 0$ VSL theories satisfy a 
weak form of the equivalence principle (and do not conflict  
with the E\"{o}tvos 
experiment); however they 
predict non-geodesic motion. Indeed the action for a point 
particle, with $b\neq 0$, is given by 
\be\label{partaction} 
S=-{E_0\over 2}\int d\lambda e^{b\psi}g_{\mu\nu}\dot x^\mu\dot x^\nu  
\ee  
in which the ``$x^0$-affine'' parameter is given by $d\lambda=c d\tau$, 
where $d\tau$ is the actual affine parameter (proper time in the VSL 
units, for a time-like particle).  
Hence, if $b\neq 0$, particles do not follow lines  
of extremal length, but instead minimize the functional (\ref{partaction}). 
Varying (\ref{partaction}) shows that source terms appear in the geodesic  
equation, specifically  
\be  
\ddot x^\mu +\Gamma^\mu_{\alpha\beta}\dot x^\alpha \dot x^\beta  
=-b{\left(\dot x^\mu\dot x^\nu-{1\over 2}g_{\alpha\beta}\dot x^\alpha  
\dot x^\beta g^{\mu\nu}\right)}\psi_{,\nu}  
\ee

Consider now radial geodesics ($\dot\theta=\dot\phi=0$) 
in the Schwarzchild metric, so that: 
\bea 
{\cal L}&=&e^{b\psi}{\left(-B\dot \xi^2   
+{\dot r^2\over B}\right)}\\ 
{c\over c_0}&=&e^{\psi}=B^{b\over 2\kappa}\\ 
B&=&1-{2Gm\over c_\infty^2 r} 
\eea  
(we have assumed the usual limit, with $b\gg a$).
There are two  conserved quantities: 
\bea 
E&=&e^{b\psi} B\dot \xi\\ 
{\cal L}&=&-\epsilon 
\eea 
(with $\epsilon=1$ for time-like particles) 
from which we derive: 
\be 
\dot r={\sqrt {E^2 B^{-{b^2\over\kappa}}- B^{1-{b^2\over 2\kappa}}}} 
\ee 
If the speed of light does not change, we have  
\be 
\tau=\int_{r_i}^{r_h} {dr\over c_0 {\sqrt {E^2 -  B}} } 
\ee 
(where $r_i$ and $r_h=2Gm/( c_\infty^2 )$ label the starting  
point and the horizon), and so the proper time taken for a free 
falling observer to reach the horizon converges.  
However, as is well known, such a process takes infinite coordinate 
time: 
\be 
t=\int_{r_i}^{r_h} {E\over c_0 B} dr=\infty 
\ee 
If $c$ changes, the proper time required to reach the horizon is now: 
\be 
\tau=\int_{r_i}^{r_h} {dr\over c  
{\sqrt {E^2 B^{-{b^2\over \kappa}}-B^{1-{b^2\over 2\kappa}}}}} 
\ee 
Let us first assume that $b\ll 1$ but $b/\kappa$ is non-negligible, 
(so that $b^2/\kappa\ll 1 $).  
Then this differs from the fixed $c$ case in that ``$v\propto c$'',  
as naively expected. Hence the horizon is unreachable if $c$ goes to  
zero faster than $r-2m$, that is if $b/(2\kappa)\ge 1$. 
When $b^2/\kappa$ is non-negligible, 
the field $\psi$ also acts as an extra gravitational force, 
accelerating or braking free-falling particles.  In the general case 
$\tau$ diverges if 
\be
{b\over 2\kappa}(1-b)\ge 1
\ee
(with $1+b^2/(2\kappa) >0$ and $\kappa >0$).
 
In general (that is without assuming a Schwarzchild solution) 
there are regions of parameter space 
for which the horizon may be regarded as a boundary of space-time,  
since it is located at infinite affine distance from any point 
in its exterior.

\subsection{Interaction clocks in the vicinity of black holes} 
\label{intclock}
However one should bear in mind an extra complication, already  
discussed in \cite{li}. Interaction paces also change near the black hole, 
since all fine structure constants change. Strong decays 
are faster than weak ones because $\alpha_s\gg \alpha_w$.  
Similarly, as the strength of all interactions varies near the black 
hole, so will the time rates of all the processes they promote. 
 
Somewhat philosophically it was pointed out in \cite{li} 
that our sensation of time flow derives precisely from change, 
and this is imparted by interactions and their rates. Hence 
we introduced the concept of an ``interaction clock'', a device 
ticking to the time scales set by the $\alpha_i$ (the fact that 
the ratios between all $\alpha_i$ are constant removes any ambiguity). 
The tick of such a clock is given by $\tau_0(\alpha_i)=\tau_0(c)$ 
\cite{li}, with  
\be \label{tau} 
\tau_0={\hbar\over \alpha^2 Q} \propto {1\over c^{2q+1}} 
\ee  
in which $Q$ is the energy scale of the process producing the 
tick $\tau_0$. 
One such construction is a muon clock. Let us produce a large 
number of non-relativistic muons. When half of them have decayed the  
clock ticks, and produces another large number of muons. Such a clock 
would tick to a rate \cite{mandl}: 
\be  
\tau_\mu={96\pi^3\hbar\over E_\mu\alpha_w^2 {\left( m_\mu\over m_W\right)}}  
\ee  
where $m_\mu$ and $m_W$ are the muon and the W masses, and  
$\alpha_w=g_w/(\hbar c)$ is the weak fine structure constant. 
Another example is an atomic clock, the period of which is 
given by 
\be 
\tau_e={\hbar\over \alpha_e^2 E_e} 
\ee 
where $E_e=m_e c^2$ is the electron rest mass. Since 
$E_e\propto c^q$ (like all other relativistic 
energies) we have that $\tau_e\propto {1/ c^{2q+1}}$.  
These are two realizations of interaction clocks; if all else 
fails remember that $\tau_0$ is the pace at which we age 
\cite{BT}. 
 
A better formulation of the question of whether an observer 
may or may not reach the horizon is then: how many $\tau_0$ ticks 
are required?  For a Schwarzchild solution 
this means computing the dimensionless number: 
\be 
{\cal N}=\int_{r_i}^{r_h} {d\tau\over \tau_0} 
=\int_{r_i}^{r_h} {dr\over \tau_0 c  
{\sqrt {E^2 B^{-b^2\over \kappa}-B^{1-{b^2\over 2\kappa}}}}} 
\ee 
which diverges if  
\be 
-{b\over 2\kappa}[2q+b]\ge 1 
\ee 
This condition defines the parameter space for which the horizon 
should be counted as part of the spatial infinity of the black 
hole.

\subsection{Are there VSL singularities?} 
This result is extremely interesting. Our solution has a  
singularity at $r=0$ (in some cases for the general solution 
there is in fact a naked singularity at $r=r_h$). However 
this singularity is physically inaccessible; not just in the sense 
that information cannot flow from it into the asymptotically flat  
region, but also in the sense that no observer starting from 
the asymptotically flat region can actually reach it. The singularity 
lies in a disconnected piece of the manifold, which should simply  
be excised as unphysical.

It is tempting to conjecture that all singularities are subject to  
the same constraint, in which case we seem to have eliminated the 
singularity problem, by means of a stronger version 
of the cosmic censorship principle.

\section{Collapsing stars and their remnants}\label{col} 
We now discuss stellar collapse making use of the  
Oppenheimer-Snyder solution, in which a spherical dust 
ball collapses. This is the correct solution  
in the limit $a,\kappa,b\rightarrow 
0$, keeping $\kappa/a$ and $\kappa/b$ finite. In this case
the metric is Schwarzchild and motion is geodesics.
Other cases are more complicated (see \cite{teuk}  
for an investigation in the context of Brans-Dicke theory). 
 
The Oppenheimer-Snyder solution makes use of Birkoff's 
theorem to match a Schwarzchild outside solution, to a Friedmann 
closed solution in collapsing stage (in general we note that the  
solutions derived in Section~\ref{bh} apply to the outside of a  
static star  
for the same reason). The inside metric is then: 
\be 
ds^2=-d\zeta^2+R^2(\zeta)[d\chi^2+\sin^2\chi d\Omega^2] 
\ee 
Here $\zeta$ is the proper $x^0$ of free-falling observers, 
$\chi$ is the radial coordinate of a 3-sphere, and $R$  
is the expansion factor. The latter  
satisfies standard Friedmann equations (which are valid 
in the regime under study \cite{vcb}) for a dust Universe 
with density $\rho$. One can show that there is no jump in the  
curvature provided that 
\bea 
m&=&{4\over 3}\pi \rho R_0^3\\ 
R_0&=&\sin\chi_0 R(\zeta) 
\eea 
in which $\chi_0$ is the radial coordinate indexing the surface 
of the star (which follows a geodesic). The internal value 
for the speed of light is given by 
\be 
\psi={b-a\over \kappa}\log {\left( 1-{8\pi G\rho 
\sin^2\chi_0R^2\over 3c_\infty^2}\right)} 
\ee 
Even though the Oppenheimer-Snyder solution may be adapted 
to our circumstances, the physics of collapse is entirely 
different. The arguments applied in the previous Section 
to free falling observers are also valid for observers 
on the surface of the star.  
In standard relativity collapse takes infinite  
coordinate time, but finite proper time for an observer  
on the surface of the star. In VSL theories the proper time, 
as felt by interaction clocks on the surface of a collapsing 
star, is infinite (for the parameter region identified in the last Section).  
As the surface of the star 
approaches its Schwarzchild radius, all processes freeze-out. 
We are left with a Schwarzchild remnant, the surface of which is part 
of spatial infinity.  The star itself has left the manifold.  
It's still black, but it's not a hole; rather its surface is 
an edge of space.  
 
When this happens there is also a divergence for the number 
of ticks for any process for observers inside the star, since 
$c$ inside the star must also go to zero or infinity.  
Hence the singularity is never formed, a fact which in any case 
has little physical relevance. 
The inside of the star is pickled for eternity as the Schwarzchild 
remnant is formed.

\section{Gravitational physics around stars}\label{phys} 
We now turn to the  
study of gravitational phenomena in the vicinity 
of VSL stars. A more detailed study within  
the framework of the PPN formalism \cite{will}  
is warranted, but shall not  be attempted here.   
 
In summary we find the following. There are 
three classes of effects: upon planetary orbits (eg. the precession 
of the perihelion of Mercury), upon light (eg. gravitational light 
bending, or the radar echo time-delay), and upon the fine structure of 
absorption lines. These are caused, in different combinations, 
by three distinct facts which 
we can switch on and off independently: corrections to the  
Schwarzchild metric, violations of energy conservation, and  
spatial variations in $\alpha$.  
 
If there are only corrections to the Schwarzchild metric 
we obtain corrections to the GR result  
for the planetary and light trajectories  
similar to those found in Brans-Dicke theory. These corrections 
are embodied in the PPN parameter $\gamma$ computed above. 
However there is a limit in which we recover the Schwarzchild metric  
(and $\gamma=1$) but in which there are significant  
violations of energy conservation. In this limit we recover the GR  
results for light properties, but we find non-negligible  corrections  
to planetary orbits. Finally it is possible to switch off 
these two effects, and so recover the classical tests of GR, 
and still produce significant changes in $\alpha$ and consequently 
in the fine structure of spectra in light emitted at the surface of 
stars. It is also possible to switch off the latter, and keep either  
of the former two effects.

\subsection{The precession of the perihelion of Mercury}  
We start by deriving the orbits of point particles, considering  
first the Schwarzchild metric. We are therefore in the limit
$a,\kappa\rightarrow 0$, but we shall assume that $b$ is finite
and $\kappa/a\gg b$ so that we may exhibit deviations from 
geodesic motion. Setting $\theta=\pi/2$, $\dot\theta=0$, 
the Lagrangian is (cf. Eqn~(\ref{partaction})):  
\be  
{\cal L}=e^{b\psi}{\left(-B\dot \xi^2   
+{\dot r^2\over B} + r^2\dot\phi^2\right)} 
\ee  
There are three conserved quantities:  
\bea 
E&=&e^{b\psi} B\dot \xi\label{3cons1}\\ 
J&=&r^2 e^{b\psi}\dot\phi\label{3cons2}\\  
{\cal L}&=&-\epsilon 
\eea 
where $\epsilon=0,1$ for light and particles respectively. 
It follows that   
\be  
\dot r^2=E^2e^{-2b\psi}-\epsilon e^{-b\psi} B 
-{J^2\over r^2}  
e^{-2b\psi}B 
\ee  
Using the standard transformations: 
\bea 
u&=&{G\over rc_\infty}\\ 
{d\over d\lambda}&=& \dot\phi {d\over d\phi} 
\eea 
and differentiating we get: 
\be 
u''+u=3mu^2 +{m\epsilon \over J^2}{\left(1+{b^2\over 2 \kappa}\right)} 
(1-2mu)^{b^2\over 2\kappa} 
\ee 
in which we have used Eqn.~(\ref{cschw}). Expanding the VSL contribution 
(terms arising from $b^2/\kappa\neq 0$) 
up to first order in $mu$ leads to: 
\be \label{uvsl} 
u''+u=3mu^2 +{m\epsilon \over J^2}{\left(1+{b^2\over 2 \kappa}\right)} 
-{m^2\epsilon \over J^2}{b^2\over \kappa}{\left(1+{b^2\over 2\kappa} 
\right)}u 
\ee  
to be compared with the Newtonian result: 
\be\label{unewton} 
u''+u={m\epsilon\over J^2} 
\ee 
and the GR result: 
\be 
u''+u=3mu^2 +{m\epsilon\over J^2} 
\ee 
The Newtonian solutions are elliptical orbits: 
\be 
u_0={m\over J^2}(1+e\cos\phi) 
\ee 
where $e$ is the eccentricity. The GR term $3mu^2$ causes 
a precession of the perihelion by  
\be 
\Delta\phi={6\pi m^2\over J^2} 
\ee 
per revolution. In the  
case of Mercury this amounts to about $43''$ per century.  
 
VSL causes two extra effects, even in the limit where the metric  
remains Schwarzchild. Firstly it causes a shift in the Keplerian mass,  
that is, the Newtonian formula still applies but with mass 
\be\label{calm} 
{\cal M}= m{\left(1+{b^2\over 2\kappa}\right)} 
\ee 
This can be guessed by comparing the relevant term in (\ref{uvsl})  
with the Newtonian expression (\ref{unewton}).  
A derivation of Kepler's third law, with a more rigorous derivation 
of (\ref{calm}) may be found in Appendix III. 
Secondly, the last term in (\ref{uvsl}) induces a shift in  
the frequency, causing a precession per revolution of: 
\be\label{delphi1} 
\Delta\phi=-{4\pi m^2\over J^2}{b^2\over 2\kappa}{\left(1+{b^2\over 2\kappa} 
\right)} 
\ee 
We see that, as announced above, 
even in the limit in which the metric remains Schwarzchild, 
VSL may induce significant corrections to the orbit of Mercury. 
 
It may make more sense to rewrite $\Delta\phi$ in terms of 
${\cal M}$, since this is the mass measured using Kepler's  
third law. Then, to first order in $b^2/ \kappa$, the joint GR  
and VSL effect is: 
\be 
\Delta\phi={6\pi {\cal M}^2\over J^2} 
{\left(1-{4\over 3}{b^2\over \kappa} 
\right)} 
\ee 
In the case of Mercury,  in addition to the usual GR effect 
there is a precession of about $57''$ times {\it minus} $b^2/\kappa$, 
purely due to violations of energy conservation.

The general case is more difficult to compute. We use the 
Eddington-Robertson form of the metric in radial coordinates: 
\bea 
ds^2&=&-B d\xi^2 + A dr^2 + r^2 d\Omega^2\\ 
B&=&1-2{M\over r}+2(1-\gamma){M^2\over r^2}\\ 
A&=&1+2\gamma{M\over r} 
\eea 
with $M$ and $\gamma$ given by (\ref{bdmass}) and 
(\ref{gamma}). Using the same techniques 
as above we arrive at: 
\be\label{u2b} 
u'^2={E^2\over ABJ^2}-{u^2\over A} -{\epsilon e^{b\psi}\over 
AJ^2} 
\ee 
When $e^{b\psi}=1$ this expression leads to standard results 
(see \cite{weinb}). Hence we should add to these results 
any corrections induced by the new terms associated with the 
$e^{b\psi}$ factor. To find the new  terms  
we need $e^{b\psi}$ up to second order in $Mu$.  
Noting that  
\be 
\rho=r{\left( 1-(1+C)u\right)} 
\ee 
and expanding (\ref{bd2}) we find: 
\be 
e^{b\psi}=1-{bC\over a}Mu +{bC\over a } 
{\left({bC\over 2a} 
-1-C\right)}(Mu)^2 
\ee 
Hence the new terms in ($\ref{u2b}$) are 
\be 
u'^2=...+{bC\over a}{Mu\over J^2}-{M^2 u^2\over J^2} 
{bC\over a}{\left(2\gamma +{bC\over 2a}-1-C 
\right)} 
\ee 
where the ellipsis denotes terms present in the fixed $c$ 
calculation for PPN metrics. This leads to 
\be 
u''=...+{bC\over 2a}{M\over J^2}-{M^2 u\over J^2} 
{bC\over a}{\left(2\gamma +{bC\over 2a}-1-C 
\right)} 
\ee 
Again the Keplerian mass receives a shift 
\be\label{calm1} 
{\cal M}=M{\left(1+{bC\over a}\right)} 
\ee 
As for the perihelion precession we should now add to the standard formula 
\be 
\Delta\phi_0={6\pi M^2\over J^2}{1+2\gamma\over 3} 
\ee 
(with $\gamma$ given by (\ref{gamma})) the extra term: 
\be 
\Delta\phi_1=-{2\pi M^2\over J^2} 
{bC\over a}{\left(2\gamma +{bC\over 2a}-1-C 
\right)} 
\ee 
This result reduces to (\ref{delphi1}) in the limit $a,\kappa\rightarrow 0$, 
and $\kappa/a\gg b$. An expression containing only physically 
meaningful quantities can then be obtained by rewriting these formulae 
in terms of ${\cal M}$ by means of (\ref{calm1}).

\subsection{Gravitational light deflection} \label{light} 
Considering now light trajectories, we should set $\epsilon=0$ 
in equation (\ref{u2b}). This cancels out the term in $e^{b\psi}$  
and so VSL induces 
no effects on light trajectories other than those induced 
by distortions to the Schwarzchild metric. Hence if $a,\kappa\rightarrow 0$, 
and $\kappa/a\gg b$ we predict the same result as GR for  
gravitational light bending: 
\be\label{light1} 
\Delta\phi={4Gm\over r_0 c_\infty^2} 
\ee 
where $r_0$ is the impact parameter. In the case of a light 
ray grazing the Sun $\Delta\phi=1.75''$. The general case is: 
\be\label{light2} 
\Delta\phi={4GM\over r_0 c_\infty^2}{1+\gamma\over 2} 
\ee 
with $\gamma$ given by (\ref{gamma}).  
 
It would seem at first that in the limit in which the metric remains 
Schwarzchild there are no corrections to GR for light bending, but 
the formula for the perihelion of Mercury precession may be modified. 
This is a distinctive feature of VSL, distinguishing it from  
Brans-Dicke theories, and can be traced to violations of energy 
conservation in the Jordan frame in these theories.  
In practice however the situation is very different. The masses 
$m$ or $M$ are not directly accessible; the mass of the Sun 
being estimated  
via Kepler's law. The result is a Keplerian mass ${\cal M}$ given by 
either (\ref{calm}) or (\ref{calm1}). Hence, even though VSL corrections 
of order $b^2/\kappa$ only affect time-like orbits, these corrections  
filter through to formulae for light trajectories, because these must be 
expressed in terms of Keplerian masses. The relevant result is 
obtained by substituting (\ref{calm}) or (\ref{calm1}) in  
(\ref{light1}) or (\ref{light2}).

This situation is a good object lesson against harsh applications of conformal 
transformations. As spelled out in Appendix II, if we ignore 
the most general type of ${\cal L}_m$ it is possible to map the  
dynamics of our theory into Brans-Dicke dynamics. This explains why 
our formulae for planets (which are not conformally 
invariant) differ from Brans-Dicke  results, but the same does not 
happen to light (which is conformally invariant). However, such a 
direct application of a conformal transformation would miss 
the interconnection between conformally invariant and non-invariant
results which we have just pointed out.


\subsection{Radar echo time-delay} 
Naively one might expect a different result for radar echo 
time delays in VSL theories. Indeed if light travelled slower/faster  
near the Sun, the echo time-delay should be larger/smaller.  
As we shall see this is not true in our theory, a feature due 
to the fact that we have not broken local 
Lorentz invariance. As pointed out 
in \cite{li} this manifests itself in the absence of a global time 
coordinate, the differential structure associated with time forming 
a fibre bundle rather than a tangent bundle. Hence non-local calculations 
involving time should be done with the coordinate $\xi$, the conversion 
to time to be done locally. As a consequence whatever happens to $c$ locally 
along the path of the radar wave does not affect the final result.  
 
We start by deriving results valid {\it if} 
we were to break local Lorentz invariance.
Let $r_0$ be the point of closest approach 
to the Sun. Then the time taken for the radar signal to move up to  
distance $r$, in the absence of gravitational effects,  is: 
\be 
\Delta t=\int dr {r\over c(r)[r^2-r_0^2]^{1/2}} 
\ee 
With a variation in $c$ analogous to (\ref{cschw}) 
we would get  
\be\label{delt1}
\Delta t= {[r^2-r_0^2]^{1/2}\over c_\infty} +{b-a\over
\kappa}{Gm\over c_\infty^3} 
\log{\left(r+ [r^2-r_0^2]^{1/2}\over r_0\right)} 
\ee 
Hence to the usual gravitational time-delay, we would have to add 
a delay (if $\alpha>0$) due to a lower value for $c$ close to the Sun.
Comparing (\ref{delt1}) with the usual PPN formula \cite{weinb}
we find that  this effect, due to  explicit violations of Lorentz 
invariance, simulates a PPN parameter $\gamma=(b-a)/\kappa$.

Nothing like that happens in a locally Lorentz
invariant  VSL theory. From: 
\be 
\dot r^2={E^2e^{-2b\psi}\over AB}-\epsilon e^{-b\psi} B 
-{J^2\over r^2}  
e^{-2b\psi}B 
\ee  
we obtain, after setting $\epsilon=0$ and making use of:
\be 
\dot r={dr\over d\xi}{E\over e^{b\psi}B} 
\ee 
the expression: 
\be 
{\left( dr\over d\xi\right)}^2= 
{1-{J^2 B\over E^2}\over A/B} 
\ee 
in which all factors in $e^{b\psi} $ have cancelled out. 
This leads to the standard expression \cite{weinb}  
\bea\label{delay} 
\Delta \xi&=& [r^2-r_0^2]^{1/2}+ 
\nonumber\\ 
&&(1+\gamma){GM\over c_\infty^2 } 
\log{\left(r+ [r^2-r_0^2]^{1/2}\over r_0\right)} +\nonumber \\ 
&&{GM\over c_\infty^2 }{\left(r-r_0\over r+r_0\right)}^{1/2} 
\eea 
One needs now to transform $\xi$ into time, but that is done 
on the Earth, where $c\approx c_\infty$. Hence   
$\Delta t=\Delta\xi/c_\infty$, leading to the same 
result as in Brans-Dicke theories. 
 
The only novelty is again that $M$ is not the Keplerian mass, 
if $b^2/\kappa$ is non-negligible. Once more we find that even though 
VSL is equivalent to Brans-Dicke  in light experiments, the fact that 
masses are estimated using time-like objects induces corrections 
in formulae for light. In the present case we should use
(\ref{calm1}) to replace  
$M$ with ${\cal M}$ in (\ref{delay}).

\subsection{Spectral lines}  
Naturally the hallmark and real novelty of VSL is  
a changing electromagnetic fine structure constant. This 
should affect the fine structure of absorption lines created on  
the surface of stars, and be detectable using techniques similar 
to Webb et al \cite{webb}. As we shall see the larger the potential  
difference, the stronger the effect, so perhaps dwarfs, or even neutron  
stars might be better candidates for this experiment.

We first consider the effect upon spectra in the non-relativistic
regime. We find that all spectral lines are proportional to
the Rydberg energy, given by 
$E_R=m_e e^4/\hbar^2=E_e\alpha^2$, where $E_e=m_e c^2$ is
the electron's rest energy. Hence spectral lines  have wavelengths 
proportional to $\lambda=\hbar c/E_R\propto1/\alpha^2\propto
c^{-2q}$. Considering that photons in free flight 
have a constant wavelength (see Section V~A of \cite{li}) we conclude
that when we compare spectral lines coming from the surface of a star
with those measured on an Earth laboratory, we find an extra
``redshift'' effect, due to VSL, of magnitude:
\be
{\Delta\lambda\over \lambda}=-2q{\Delta c\over c}=
{2bq\over \kappa}{G m\over c_\infty^2 r} 
\ee
where the last identity is valid only in the limit 
$a,\kappa\rightarrow 0$, and $\kappa/a\gg b$.
We therefore conclude that VSL theories have a PPN 
parameter $\alpha_{PPN}=2bq/\kappa$ \cite{will}.
Pound-Rebka-Snider experiments are capable of constraining
this  parameter, but not by more than $|\alpha_{PPN}|
<10^{-3}$ (see Fig.14.3 of \cite{will}). As will be shown in
\cite{vcb} the combination $bq/\kappa$ is of order
$\Delta\alpha/\alpha$ at cosmological redshifts or order 1.
Hence the observations made by Webb et al \cite{webb}, when
interpreted with VSL, imply violations of the weak equivalence
principle at the level $\alpha_{PPN}\sim 10^{-5}$, consistent
with current experimental tests. In particular, measurements
of non-relativistic spectral lines formed on the surface of
the Sun do not constrain $\alpha_{PPN}$ by more than 
$|\alpha_{PPN}|<10^{-2}$. More compact objects, such as
dwarfs or pulsars, display a stronger VSL redshift effect,
but the effect is degenerate with respect to 
Doppler shifts induced by their unknown velocities with respect to us. 
For such objects one has to go to look
into the fine structure in order to measure, without degeneracy,
the possible effects upon spectra lines of varying constants.

Considering now the relativistic fine structure of spectral lines, we find
that they directly measure the tell-tale 
signature of VSL, since they are directly related to
$\alpha=e^2/(\hbar c)$ (not to be confused with $\alpha_{PPN}$). 
For small deviations we have: 
\be  \label{dela}
{\Delta\alpha\over \alpha}=q{\Delta c\over c}=q\psi 
=-{q(\gamma -1)\over a}{G M\over c_\infty^2 r} 
\ee 
where we have used (\ref{bd2}) (recall that $\alpha\propto c^q$).  
It is interesting to note that (\ref{dela}) may be large 
even choosing parameters which render the metric Schwarzchild, 
and non-geodesic effects associated with $b\neq 0$ negligible. 
In the limit
$a,\kappa\rightarrow 0$, and $\kappa/a\gg b$ (so that 
$\gamma\approx 1$) we have 
\be 
{\Delta\alpha\over \alpha}=-{bq\over \kappa}{G m\over c_\infty^2 r} 
\ee 
This may be non-negligible even with negligible $b^2/\kappa$ (so that 
no corrections to the GR result are present in the perihelion of Mercury). 
The prefactor $bq/ \kappa$ may be inferred from cosmological
observations \cite{vcb}
and can at most be of order $10^{-4}$. Hence we need
an object sufficiently compact, such as an AGN, a pulsar or a white
dwarf, for the effect to be non-negligible. Furthermore we need
the ``chemistry'' of such an object to be sufficiently simple,
so that line blending does not become problematic
\footnote{ I would like to thank Lance Miller and
Gra\c{c}a Rocha for tutoring me on the details of stellar
spectral lines.}.
 
Generally (i.e. for any matter configurations) the larger the 
gravitational potential differences, the stronger the effect. 
Indeed, for static configurations, both $\Delta\alpha/\alpha$ and the  
gravitational potential satisfy Poisson equations, with
source terms related by a multiplicative constant. Hence the local 
value of $\alpha$ should map the  
gravitational potential, and one would need to have big variations 
in the gravitational potential to observe corresponding spatial  
variations in $\alpha$. It would be interesting to use this 
to infer $\alpha$ maps from $N$-body simulations, so as to deduce
possible observational signatures of VSL on cluster and supercluster
scale.


\section{Theoretical and observational outlook} 
We have provided ample evidence for how VSL stars and ``black holes''
may be rather exotic indeed. We have used the covariant and locally
Lorentz invariant formulation proposed in \cite{li}, and stress that
the results derived are by no means generic to all VSL theories. Indeed
in Appendix I we showed how bimetric VSL black holes may differ distinctly
from the ones considered here. In this regard 
it would be of great interest to
derive the properties of black holes in the bimetric theory of
Clayton and Moffat \cite{cly1,cly2,cly3} and Drummond \cite{drum}.
Another variation upon the theme are VSL theories which explicitly 
break local 
Lorentz invariance, such as the one proposed by Albrecht and Magueijo
\cite{am}, and for which black hole solutions remain elusive. 
In Section VII C we derived a distinctive
effect to be expected in such theories (a different radio echo
time-delay) which is not present in locally Lorentz invariant
VSL theories. Hence the exotic results derived in this paper
are generic to the type of theories proposed in \cite{li}, but
by no means to all VSL theories.

Yet, even within the framework of the VSL theories proposed in \cite{li},
 a large number of new effects still remain to be explored. We close 
this paper firstly by  
highlighting a few obvious areas of interest which should prompt
further theoretical work, and then describing observational prospects.

An important omission in this paper is quantum effects, 
which we have ignored. However it was
shown in \cite{li} that a varying-$c$ induces quantum particle creation 
(a point noted before, in other VSL theories, by \cite{harko}).
That being the case, VSL black holes might be sources of radiation
in a process complementary to Hawking's radiation. The exact details
of such a process remain to be worked out. Also the interaction 
between a changing $c$  and standard Hawking radiation is far from obvious.
These phenomena are currently being investigated. 

Further quantum effects arise from the fact that all gauge field strengths 
becoming zero or infinite will no doubt reshape the low-energy aspect of 
any quantum field theory.
Indeed, the scaling arguments mentioned in Section~\ref{intclock}
should break down when the line $\alpha=1$ is crossed. Therein 
non-perturbative interactions will become  perturbative, or vice-versa,
a process which may have dramatic implications. For instance,
the vacuum of a given
theory may change. The impact upon phenomena like confinement may
be massive.

There are also other interesting classical effects beyond those
 described in this paper. All the arguments
developed in this paper concerned free-falling point particles.
One may wonder what happens to free-falling extended objects. As
is well known, they will feel gravity by means of tidal forces.
Should $b\neq 0$ they will also feel inertial forces, corresponding
to their acceleration (or braking) by the field $\psi$. Furthermore
there will also be effects induced by the gradients in $c$. 
Let us consider a body moving along a negative gradient 
of $c$ (and assume  $b=0$). Given that $v\propto c$, such a body  would 
get squashed along the direction of  motion. In general a stress
proportional to ${\bf v}\cdot \nabla \psi$ will be felt.

Another finite size effect involves the time rates associated
with ``interaction clocks'' derived in Section~\ref{intclock}. 
For a point particle falling into a black hole a slowing down
of this rate means merely the slowing down of its progression 
towards the horizon. However, for an extended object there
will also be an aging gradient, closely 
mapping the $c$ gradient, in addition to the stresses mentioned
above. These issues, as well as the quantum effects described above,
will be the subject of a future publication.

Besides these interesting topics for future theoretical
work, there is the obvious hurdle of experiment.
We saw that the theory produces effects very 
similar to Brans-Dicke theory, plus additional effects, namely 
departures from geodesic motion for non-null particles, and 
distorted fine structure in spectral lines in stellar light. 
If $b=0$ the classical tests of GR impose the constraint
\cite{will}:
\be
|\gamma-1|<10^{-3}
\ee
If we adopt the Schwarzchild limit (in which case $\gamma=1$)
this constraint becomes 
\be
{b^2\over |\kappa|}<10^{-3}
\ee
In between these two limits a rather complex combination of
$a$, $b$, and $\kappa$ is constrained to the same order of
magnitude. 

Should there be any departures from GR results in these classical
experiments, however, VSL would be an interesting competitor
to Brans-Dicke theory, since it predicts corrections to light and
planetary formulae distinct from Brans-Dicke theory. More interesting
still is that, unlike Brans-Dicke theory, the theory does not
become trivial in the limit in which the classical tests of GR
are reproduced ($\omega_{BD}\gg 1$ for Brans-Dicke, 
$a,\kappa\ll 1$, $\kappa/a\gg b$, and $b^2/\kappa\ll 1$
for VSL). In this limit the theory still predicts a shift in $\alpha$
observable in the fine structure of spectra from stars or
other compact objects. This
effect makes VSL an interesting experimental target.

\section*{Acknowledgements}I would like to thank 
Andy Albrecht, John Barrow, Kim Baskerville, Malcolm Perry,
Kelly Stelle, and John Webb,  for discussion or comments. I am grateful to  
the Isaac Newton Institute for support and hospitality while 
part of this work was done.

\section*{Appendix I - Black holes in bimetric theories} 
As a curiosity we now show an example of an alternative
VSL theory which evades the theorem described in  
Section~\ref{proof}. We show how this happens, 
using as an example the theory described in the Appendix 
of \cite{li}. In this theory there are two metrics, $g$ coupling  
to gravitation and matter, and  $h$ coupling to the field $c$ only.  
The action is:   
\bea\label{S3}     
S&=& S_1+S_2\nonumber\\  
S_1&=&  
\int d^4x \sqrt{-g}{\left(   
R+{ 16\pi G\over c_0^4e^{4\psi}}{\cal L}_m \right)} \nonumber\\  
S_2&=&  
\int d^4 x\sqrt{-h}{\left( H 
-\kappa h^{\mu\nu}\partial_\mu\psi\partial_\nu\psi  \right)}  
\eea   
where $g_{\mu\nu}$ and $h_{\mu\nu}$ lead to two Einstein tensors  
$G_{\mu\nu}$ and $H_{\mu\nu}$. 
Varying with respect to $g$, $\psi$, and $h$ leads to equations:  
\bea   
G_{\mu\nu}&=&{8\pi G\over c_0^4e^{4\psi}}     
T_{\mu\nu}\\   
\Box_h \psi&=&{32\pi G\over c_0^4e^{4\psi}\kappa}\sqrt{g\over h}   
{\cal L}_m\\   
H_{\mu\nu} &= & \kappa    
{\left(\nabla_\mu\psi \nabla_\nu\psi     
-{1\over 2}h_{\mu\nu}\nabla_\alpha\psi \nabla^\alpha\psi     
\right)}     
\eea   

Let us now consider SSS solutions to this theory. 
It is immediately obvious that $g_{\mu\nu}$ is the Schwarzchild
solution, with mass $m$. The solutions for $c$ and $h_{\mu}$
can be obtained by applying to this theory an argument
similar to the one followed in Section~\ref{bhbd}. Solutions
(\ref{bd1}) and (\ref{bd2}) are still valid, since we are
in vacuum. However the weak field limit now produces:
\bea
-h_{00}&=&1-{2Gm\over c_\infty r}{-ab\over 2\kappa}\\
\psi&=&-{b\over \kappa}{Gm\over c_\infty r}
\eea
in which we have $b=-4$ and $a\rightarrow 0$. Hence,
comparing with the asymptotic forms of (\ref{bd1}) and (\ref{bd2}),
we find
\bea
C&=&-2\\
\lambda a&=&{\sqrt{2\kappa}}
\eea
This leads to the result for $h_{\mu\nu}$ and $c$:
\bea 
ds^2&=&-d\xi^2 
+{\left(1-{\left(\rho_0\over \rho\right)}^2\right)}^2  
(d\rho^2 +\rho^2 d\Omega^2)\\
c&=&c_0{\left(1-\rho_0/\rho\over 1+\rho_0/\rho\right)}
^{-{\sqrt {2/\kappa}}}
\eea 
in which the ``horizon'' is at
\be
\rho_0={\sqrt{2} m\over \kappa}
\ee
The horizon of $g_{\mu\nu}$ and that of $h_{\mu\nu}$ (which is where
$c$ goes to infinity) therefore do not need to be at the
same place.

\section*{Appendix II - Conformal duals} 
Here we examine the effect of conformal transformations 
on VSL theories. These are to be distinguished 
from changes of units which render $c$ constant, leading to fixed 
$c$ duals, as studied in \cite{li}. Conformal transformations 
take the form 
\bea  
{d\hat t}&=&dt \Omega\\  
{d\hat x}&=&dx \Omega\\  
{\hat g}_{\mu\nu}&=& g_{\mu\nu}\\ 
{d\hat E}&=&dE \Omega^{-1} 
\eea  
or equivalently 
\bea  
{d\hat t}&=&dt \\  
{d\hat x}&=&dx \\  
{\hat g}_{\mu\nu}&=& \Omega^2 g_{\mu\nu}\\ 
{d\hat E}&=&dE \Omega^{-1} 
\eea  
These transformations do not change the value of $c$, and so map 
VSL theories into VSL theories;  but the gravitational action is 
modified leading to different values for $a$, $b$, and $\kappa$.  
The point we wish to make is that the degeneracy of conformally 
related theories is usually broken by the presence of matter. 
Indeed minimal coupling (the requirement that ${\cal L}_m$ does 
not depend on $c$) is not conformally invariant, and so a 
conformal frame (and so a set of $a$ and $b$) is picked for 
its simplicity in describing non-gravitational physics (a point 
clearly made in \cite{qui}). Another example of a case where 
a preferred ``physical'' conformal frame is present was given 
in \cite{Dam90}. 
 
If we can ignore generic matter fields, however, conformal 
transformation may be a useful mathematical trick. Of particular  
interest is the ``Jordan'' or ``geodesic'' frame, 
in which $b=0$, and $\psi$ does not couple to ${\cal L}_m$. 
In such a frame there is energy conservation, and particles follow 
geodesics. Two other frames of interest are the Einstein frame 
($a=0$) and the string frame ($a=b$). 
 
Consider then an action of the form: 
\bea\label{action} 
S&=& \int d^4 x \sqrt{- g}(\phi^\alpha 
{\hat R}-{\omega\over \phi^\beta}\nabla_\mu\phi \nabla^\mu 
\phi - V(\phi) \nonumber\\ 
&&+{ 16\pi G_0\over c_0^4}f(\phi){{\cal L}}_m )  
\eea  
in which, in our case, $f(\phi)=\phi^{b/a}$. Under 
a conformal transformation ${\hat g}_{\mu\nu} 
=\Omega^2 g_{\mu\nu}$, the transformed action is: 
\bea 
{\hat S}&=&{\int d^4 x} \sqrt{- {\hat g}}\{ \Omega^{-2}\phi^\alpha 
{\hat R}+6\phi^\alpha\Omega^{-4}{\hat \nabla}_\mu \Omega 
{\hat \nabla}^\mu \Omega\nonumber\\ 
&&-6\alpha\phi^{\alpha-1}\Omega^{-3}{\hat \nabla}_\mu \phi 
{\hat \nabla}^\mu \Omega 
-\omega \phi^{-\beta}\Omega^{-2}{\hat \nabla}_\mu \phi 
{\hat \nabla}^\mu \phi \nonumber\\ 
&&- V(\phi)\Omega^{-4} +{ 16\pi G_0\over c_0^4}f(\phi){\hat{\cal L}}_m  
({\hat g}_{\mu\nu}\Omega^{-2})\} 
\eea  
If a portion of ${\cal L}_m$ is homogeneous  
degree $\alpha$ in the metric, it is possible to transform away any  
coupling between $\phi$ and  ${\cal L}_m$ by setting $\Omega^2= 
\phi^n$ with $n=b/(a\alpha)$. Note that this is only possible if 
${\cal L}_m$ is homogeneous in the metric, something which is not 
generally true (for instance kinetic terms are first order in the 
metric whereas interaction terms are zeroth order). If we stick 
to classical particles, $\alpha$ is the power 
of $u$ to be used in the Lagrangian: 
\be  
S=-{E_0\over 2\alpha}\int d\lambda [g_{\mu\nu}\dot x^\mu\dot x^\nu]^\alpha 
\ee  
In standard GR this does not matter, but here it is crucial. In 
\cite{li} we have argued for $\alpha=1$, but this need not be 
the case ($\alpha=1/2$ is the value usually used in the literature, 
so that the action becomes the length of the world-line). We 
could even consider the case in which different types of classical 
matter had different $\alpha$, another good example of a situation 
in which the geodesic frame would not exist (as indeed  ${\cal L}_m$ 
would then not be homogeneous in the metric). 
 
Action (\ref{action}) then becomes: 
\bea 
{\hat S}&=&{\int d^4 x} \sqrt{- {\hat g}}\{ \phi^{1-n} 
{\hat R}-{\omega +3n(1-n/2)\over  \phi^{1+n}} 
{\hat \nabla}_\mu \phi {\hat \nabla}^\mu \phi \nonumber\\ 
&&- V(\phi)\phi^{-2n} +{ 16\pi G_0\over c_0^4}f(\phi){\hat{\cal L}}_m  
({\hat g}_{\mu\nu}\Omega^{-2})\} 
\eea 
Setting: 
\be 
\chi=\phi^{1-n} 
\ee 
and $V=0$ we finally recover the Brans-Dicke action with: 
\be 
{\hat S}={\int d^4 x} \sqrt{- {\hat g}}\{ \chi 
{\hat R}-{{\hat \omega} \over  \chi}{\hat \nabla}_\mu \chi  
{\hat \nabla}^\mu \chi 
+{ 16\pi G_0\over c_0^4}{\hat{\cal L}}_m ({\hat g}_{\mu\nu})\} 
\ee 
with  
\be 
{\hat \omega}={\omega+3n(1-n/2)\over (1-n)^2} 
\ee 
 
By means of this transformation it is now possible to  
confirm most of the results derived in this paper.  
For instance, Eqn.~\ref{gamma} may be derived from the 
usual Brans-Dicke result (with terms in $b$ included). 
On the contrary the careless application of this tool to the  
prediction of the precession of the perihelion of 
Mercury and gravitational light deflection may be very 
misleading. One might expect light properties to remain 
unaffected by this transformation. While this is true on the surface, 
it is not in reality. Formulae for the light deflection  
contain the Keplerian mass, which is affected by conformal transformations. 
This point is made clear in Section~\ref{light}.

 
\section*{Appendix III - Keplerian orbits in VSL theories} 
The Keplerian mass is estimated from Kepler's third law, which 
here we simplify to circular orbits. Then planets at distance 
$R$ have periods $T$ such that $R^3/T^2$ is a constant, proportional 
to the mass of the Sun. Kepler's law is used to estimate the mass 
of the Sun, and therefore any corrections it receives filter through 
to all formulae involving the mass of the Sun. 
 
We consider first a VSL Schwarzchild metric, so that (cf.~(\ref{u2b})): 
\be 
u'^2={E^2\over J^2}-u^2 B - {e^{b\psi}B\over J^2} 
\ee 
Following \cite{weinb} we now set to zero both $u'$ and also  
its derivative with respect to $u$ (the latter required for  
stability of the orbit). This leads to  
\bea 
E^2&=&B(J^2 u^2 + e^{b\psi})\\ 
J^2&=&{(1+b^2/(2\kappa))B^{b^2/2\kappa}B'\over u^2 
(2Bu -B')} 
\eea 
From (\ref{3cons1}) and (\ref{3cons2}) we have 
\be 
{d\phi\over d\xi}={Ju^2 B\over E}\approx 
{\left(m(1+b^2/(2\kappa)) u^3\right)}^{1/2} 
\ee 
in which the last approximation reflects the fact that  
for all planets used to estimate the mass of the Sun  
$mu\ll 1$. Hence, with $\omega=2\pi/T$, we have 
\be 
\omega^2 R^3={\cal M}=m{\left (1+{b^2\over 2\kappa}\right)} 
\ee 
 
A similar exercise using the general form of the equations
of motion confirms Eqn.~(\ref{calm1}).
 
\end{document}